\documentclass[a4paper,11pt]{article}
\usepackage{pos}
 \renewcommand{\logo}{\relax}
\usepackage{amsmath,siunitx}
\usepackage{natbib}
\title{\textit{Euclid} legacy science prospects}
\ShortTitle{\textit{Euclid} legacy}

\author*[a,b,c]{Jenny G. Sorce}
\author[d,e]{Antonino Troja}
\author[f,g]{Isaac Tutusaus}
\author[]{on behalf of the Euclid Consortium}

\affiliation[a]{Univ. Lille, CNRS, Centrale Lille, UMR 9189 CRIStAL,\\ F-59000 Lille, France\\}
\affiliation[b]{Universit\'e Paris-Saclay, CNRS, Institut d'Astrophysique Spatiale, \\91405, Orsay, France\\}
\affiliation[c]{Leibniz-Institut f\"{u}r Astrophysik (AIP), \\An der Sternwarte 16, D-14482 Potsdam, Germany\\}
\affiliation[d]{INFN-PD, \\Via Marzolo 8, Padova, Italy\\}
\affiliation[e]{Universit\'a degli Studi di Padova, \\Via Marzolo 8, Padova, Italy\\}
\affiliation[f]{Institut de Recherche en Astrophysique et Plan\'etologie (IRAP), Universit\'e de Toulouse, CNRS, UPS, CNES,\\ 
14 Av. Edouard Belin, F-31400 Toulouse, France\\}
\affiliation[g]{Universit\'e de Gen\`eve, D\'epartement de Physique Th\'eorique and Centre for Astroparticle Physics,\\ 
24 quai Ernest-Ansermet, CH-1211 Gen\`eve 4, Switzerland\\}

\emailAdd{jenny.sorce@univ-lille.fr}
\emailAdd{antonino.troja@pd.infn.it}
\emailAdd{isaac.tutusaus@irap.omp.eu}

\abstract{With the immense number of images, data, and sources that \textit{Euclid} will deliver, the consortium will be in a unique position to create/provide/construct legacy catalogues. The latter will have exquisite imaging quality and good near-infrared spectroscopy, with impact on many areas of galaxy science. These proceedings review the prospects and scientific output that \textit{Euclid} will be able to achieve in areas of galaxy and active galactic nucleus (AGN) evolution, the local and primeval Universe, studies of the Milky Way and stellar populations, supernovae (SN) and transients, Solar System objects, exoplanets, strong lensing and galaxy clusters.}

\FullConference{%
  41st International Conference on High Energy physics - ICHEP2022\\
  6-13 July, 2022\\
  Bologna, Italy
}

\begin{document}
\maketitle

\section{Introduction: why legacy?}

\textit{Euclid} is a European Space Agency-led mission designed to constrain dark energy and gravity properties using two complementary cosmological probes: weak gravitational lensing \& galaxy clustering~\citep[][]{2011arXiv1110.3193L}. To capture signatures of the Universe expansion rate and cosmic structure growth, it will cover $\sim$15\,000 deg$^2$ (Wide Survey) and 40 deg$^2$ (Deep Survey) of sky in six years in visible and near-infrared bands \citep{2014IAUS..306..375S}. It will provide images (optical down to 24.5~mag, broad-single bandpass 560--900~nm: \textit{r, i, z}, and near-infrared down to 24~mag: \textit{Y, J}, and \textit{H}) with \ang{;;0}\hspace{-0.1cm}.2 and \ang{;;0}\hspace{-0.1cm}.3 spatial resolutions respectively as well as spectra via slitless spectroscopy outside the ecliptic and Galactic planes.

As the companion proceedings of two others focusing on the two \textit{Euclid} instruments \citep[VIS and NISP,][]{2022troja} and on its main science goal \citep[cosmology,][]{2022tutusaus}, these proceedings deal only with \textit{Euclid} legacy science prospects. More precisely, legacy refers to scientific projects not necessarily designed beforehand but feasible with the data and aiming at science different from the core survey science. Such additional projects constitute an incredible added value to a given survey. The Sloan Digital Sky Survey \citep[SDSS,][]{2002AJ....123..485S} constitutes one such example with 80\% of the total official papers of the consortium that are legacy science. Moreover, since the early data release in 2002, more than 10\,000 papers have used the SDSS data. By comparison, \textit{Euclid} will provide roughly as many spectra as SDSS did in total but quasi per bin of 0.1 in redshift, in a volume almost 70 times larger, at redshifts about 1 to 3. \textit{Euclid} will thus generate images for $\sim$1.5 billion galaxies and spectra of $\sim$25 million galaxies. 
To reach the same depth, such a near-infrared imaging survey would take many hundreds of years using ground-based telescopes. These numbers reveal all the legacy potential of the \textit{Euclid} mission: the resulting large samples and volumes will permit deriving distribution functions and scaling relations, probing the extremes and finding rare sources while having also a reduced cosmic variance. Other cosmology requirements are equally useful for legacy science: the exquisite imaging will permit conducting morphological, merger and strong lensing studies. Weak lensing will provide halo properties that legacy science will link to the evolution of their hosted galaxy as well as to galaxy alignment. Spectroscopy will permit estimating galaxy metallicity and star-formation rate at $z>1$. \textit{Euclid} will touch upon many different aspects and scales of astrophysics, from the detection of brown dwarfs to galaxy formation and evolution. 

These proceedings open on a section that describes the different legacy science working groups (SWGs) currently existing within the Euclid Consortium (EC). It splits them into Galactic and extragalactic sciences. Before concluding, the complementarity with other missions is evoked. 


\section{Legacy science working groups in the Euclid Consortium}

There are eleven legacy SWGs in the EC. Two of these are discussed by the companion proceedings  \citep{2022tutusaus}: CMB cross-correlations \citep{2022A&A...657A..91E} \& simulations (Flagship). Others are detailed hereafter.

\subsection{Galactic Science}

\noindent \textbf{\textit{1. Solar System}}: Even though \textit{Euclid} points outside the ecliptic plane,   it should observe up to 150\,000 asteroids via streaks in the images.  The SWG will aim at providing astrometry, multi-color and time-resolved photometry as well as spectral classification of these Solar System objects with automated methods to find them down to kilometer sizes. The near-infrared band will provide essential details on their compositional modelling.  Combined with Gaia and LSST sparse photometry, the long lightcurves ($\sim$70~min) of \textit{Euclid} photometry will permit determining the rotation period, spin orientation, and 3D shape of the objects. Its sharp angular resolution will permit resolving binary systems in the Kuiper Belt and detecting activity around Centaurs \citep{2019EPSC...13..512C}. \\

 \noindent \textbf{\textit{2. Exoplanets / Extra-solar planets}}: Gravitational microlensing, especially from space because of a substantial increase in sensitivity to resolve main sequence stars, is a formidable tool to explore the population of cool low-mass objects such as isolated planetary mass objects, called free-floating planets. \textit{Euclid} large-scale, regular, space-based and homogeneous photometric observations will thus permit constraining the abundances of cold exoplanets down to Earth mass/size in the habitable zones of stars out to the free-floating regime. These will be compared with theoretical model predictions. Planet demographics indeed put important constraints on planet formation and evolution as well as on the frequency of violent rearrangements leading to planetary ejections \citep{2019BAAS...51c.563P}. \\

\noindent  \textbf{\textit{3. Milky Way and Resolved Stellar Populations}}: \textit{Euclid} will permit resolving individual stars in galaxies out to 7~Mpc. It will thus allow deducing accretion histories and dark matter distributions via the observations of streams and substructures. Galaxy peripheries will be mapped with stars and globular clusters. Star-formation history and chemical evolution will be derived with stars from the giant branch. New dwarf satellites will be found. It will also permit identifying low mass, stellar exotic and rare objects, probing the initial mass function and reaching ultra-cool dwarfs and the sub-dwarf boundary \citep{2020sea..confE.157M}.

\subsection{Extragalactic Science}

 \noindent \textbf{\textit{4. Local Universe -- $z<0.1$}}: \textit{Euclid} unprecedented combination of large survey area, high spatial resolution, low sky-background, and high depth makes it an excellent space observatory to probe the low surface brightness Universe limits. Stellar-halo structure and assembly will be obtained via intracluster light, tidal tails, shells, and  detections of ultra-diffuse galaxies and of globular clusters with unprecedented statistics. Their studies will provide essential information about the Universe past evolution and strong tests for the standard cosmological model  \citep{2022A&A...657A..92E}. \textit{Euclid} will hence become a cornerstone of low surface brightness astronomy for the next decades.\\

\noindent \textbf{\textit{5. Supernovae \& transients}}: Pair-instability supernovae (Pi-SN) have not yet been observationally confirmed. They are predicted in low-metallicity environments as thermonuclear explosions of very massive stars with a heavy helium core that leave behind no stellar remnants. They are useful to constrain binary black-hole formation mechanism among a large number of scenarios. Because metallicity tends to be lower at higher redshifts, \textit{Euclid} regular deep near-infrared field observations, probing high-$z$ SN, are suitable to discover such Pi-SN. Even with \textit{Euclid} low observational cadence, Pi-SN predicted long duration and high event rates should permit \textit{Euclid} to observe the first and up to several hundred Pi-SN at $z\sim3.5$. \textit{Euclid} will also discover hundreds of superluminous SN \citep{2022arXiv220409402T}.\\

\noindent \textbf{\textit{6. Galaxy and AGN evolution}}: One outstanding issue in astronomy is `understanding the physical processes that drive galaxy evolution'. While it is indeed consensual that galaxies assemble through star-formation and mergers across cosmic time as well as evolve from irregular to regular systems at lower redshifts, the exact details of this process are unknown. \textit{Euclid} will give access to billions of galaxy properties such as redshifts, stellar masses, star-formation rates to study galaxies at a critical time in the Universe star-formation history \citep{2016A&A...590A...3P}. \\
 
\noindent \textbf{\textit{7. Strong lensing}}: Strong gravitational lensing permits measuring galaxy, galaxy group and cluster total masses to a few percent. It is also sensitive to the mass slope at lensed source image positions. When combined with spectroscopy, strong lensing has a constraining power on the initial mass function of lensing galaxies. As a natural telescope, it magnifies distant source images otherwise inaccessible. About 200\,000 galaxy-scale strong lenses and 5000 galaxy clusters with lensing signatures such as arcs or bright multiple images should be found with \textit{Euclid}. These will consist in a unique laboratory to study galaxy evolution in view of the dark and luminous matter interplay.\\

\noindent \textbf{\textit{8. Clusters of galaxies}}: Galaxy clusters constitute an independent cosmological probe as long as the observable-mass relation is properly calibrated to derive accurate cluster number count in bins of mass and redshift. Clusters need thus to be blindly, but efficiently, detected with a well-known selection function \citep{2019A&A...627A..23E}. Because \textit{Euclid} will give access to galaxy clusters beyond redshift 1.5, the Figure of Merit (FoM) of the cluster number counts will drastically increase. When combined with \textit{Euclid} primary probes, galaxy clusters will permit tripling the FoM on cosmological parameters thus promising exciting results  \citep{2016MNRAS.459.1764S}.\\

\noindent \textbf{\textit{9. Primeval Universe}}: Galaxies beyond $z>6$ provide information of key importance to constrain the early phases of galaxy evolution and formation. While the number density of low-luminosity galaxies at $z>6$ is relatively high, that of bright galaxies is much lower. \textit{Euclid} wide and deep surveys at optical and near-infrared wavelengths are thus optimal to search for such objects. Thousands of $z>6$ galaxies are expected to be found, permitting studies with unprecedented statistics \citep{2022arXiv220502871E}. Additionally, rare quasars at $z>7$ may be selected from \textit{Euclid} photometry thus giving invaluable insights into conditions at the epoch of reionization when the Lyman-$\alpha$ Forest saturates and ceases to be a useful probe of reionization. Indeed, the intergalactic medium red wing is dampened when the Universe is substantially neutral \citep{2019A&A...631A..85E}.


\section{Other missions and Complementarity}

\textit{Euclid} will also greatly complement and be complemented by other missions/instruments/ob-servatories/surveys \citep[see][for details]{2017ApJS..233...21R}.  For instance, simultaneous observations by Rubin Observatory and \textit{Euclid} will not only provide independent distance measurements of Solar System objects but also improve their orbit determination. There will be indeed a significant parallax between ground-based and space Sun-Earth L2 Lagrangian point \citep[0.01~AU,][]{2018arXiv181200607S}. Similarly, simultaneous exoplanet observations by Roman and \textit{Euclid} will permit, given their large enough intra-L2 separation, microlensing parallax measurements to directly measure free-floating planet masses as well as to improve on their distance estimates even for the shortest events \citep{2019BAAS...51c.563P}. Moreover, \textit{Euclid} will provide unprecedented uniform stellar mass sampling and H$\alpha$  emitters as a function of redshift for 4MOST, WEAVE, and MOONS targeting. It will also benefit from these missions that will provide metallicity and UV properties of H$\alpha$ emitters otherwise non-accessible. Since \textit{Euclid} is slitless, these instruments will also supply better quality spectra for galaxy clusters thus improving on the dynamics knowledge \citep[vs. statistics,][]{2019most.confE..22B} as well as spectroscopic redshift sanity checks for photometric ones. \textit{Euclid} will also map a large number of cluster gravitational lenses that can feed for instance JWST. The Square Kilometer Array will permit removing H$\alpha$ emitters (last interlopers) from the Lyman-$\alpha$ lensed population at $z\geq6.6$ \citep{2017MNRAS.470.5007M}. The $z$-band data of Pan-STARRS and LSST surveys will permit refining quasar selections in the redshift interval $7<z<8$ \citep{2019A&A...631A..85E}. 

\section{Conclusion}

\textit{Euclid}, as a super SDSS for the 1 to 3 redshift Universe, will provide a data set of incredible legacy value. It will provide unprecedented uniformity of data and exquisite imaging over $\sim$15\,000~deg$^2$ in optical and near-infrared. It will thus permit conducting numerous legacy science projects from the Solar System to high-$z$ galaxies. In addition, it will nicely complement as well as be complemented by other ground-based and space missions.  In an era of precision and accuracy, joint survey processing constitutes indeed the ideal path forward. Scientifically validated data will be released to the community via the \textit{Euclid} legacy archive. In addition, quick-release data will be available for transient products except for the core cosmology objectives. \textit{Euclid} will become the reference survey of the sky, outside the galactic and ecliptic planes, for the years to come.

\begin{small} \acknowledgments

JS acknowledges support from the French Agence Nationale de la Recherche for the LOCALIZATION project under grant agreements ANR-21-CE31-0019. 
IT acknowledges funding from the Swiss National Science Foundation and from the European Research Council (ERC) under the European Union's Horizon 2020 research and innovation program (Grant agreement No. 863929; project title ``Testing the law of gravity with novel large-scale structure observables'').
The Euclid Consortium acknowledges the European Space Agency and a number of agencies and institutes that have supported the development of \textit{Euclid}, in particular the Academy of Finland, the Agenzia Spaziale Italiana, the Belgian Science Policy, the Canadian Euclid Consortium, the French Centre National d'Etudes Spatiales, the Deutsches Zentrum f\"ur Luft- und Raumfahrt, the Danish Space Research Institute, the Funda\c c\~ao para a Ci\^encia e a Tecnologia, the Ministerio de Ciencia e Innovaci\'on, the National Aeronautics and Space Administration, the National Astronomical Observatory of Japan, the Netherlandse Onderzoekschool Voor Astronomie, the Norwegian Space Agency, the Romanian Space Agency, the State Secretariat for Education, Research and Innovation (SERI) at the Swiss Space Office (SSO), and the United Kingdom Space Agency. A complete and detailed list is available on the \textit{Euclid} web site (http://www.euclid-ec.org).

\end{small}

\bibliographystyle{JHEP}
\bibliography{biblicompletenew}

\end{document}